# The effect of the toplolgy adaptation on search performance in overlay network

Muntasir Al-Asfoor[1], *Mohammed Hamzah Abed[1]

[1]University of Al-Qadisiyah, College of Computer Science and IT,

Diwaniya, Iraq

muntasir.al-asfoor@qu.edu.iq
*mohammed.abed@qu.edu.iq

**Abstract.** The work presented in this research paper has focused on the effect of network topology adaptation on search performance in peer to peer overlay network. Guided search vs. blind search have been studied with the aim of improving the search results and decreasing the time a search message would take to reach the destination. The network has been formulated as a bi-direction graph with vertices represent network nodes and edges represent connections. The level of network subject of this study is on application layer, that means two nodes are connected if they know each other contact addresses. A good example of this kind of network is the social network where all the lower layers are hidden from the end user. Two different search algorithms have been studied under these circumstances, namely: depth first algorithm and breadth first algorithm. Furthermore, the algorithms performance is examined under random topology (scale free network topology) and under topology adaptation. A simulation scenario has been designed to investigate the fidelity of the system and study the suggested solutions. Simulation results have shown that the search algorithms are performing better under topology adaptation in terms of results quality and search time.

## 1    Introduction

You Nowadays, computer networking has experienced many challenges, one of them is searching across a network with a restricted condition of reaching the destination node within a deadline. With small networks, finding the destination is rather easy and could be done subjectively. However, when a network grows large, finding the best solution is a hard job to do subjectively. Traversing a large network looking for complete solution is a challenging task in a restricted environment (time and cost).Accordingly, researchers in academia and industry have employed search algorithms, which are usually applicable for graphs, on networks' applications. By modeling the network as a graph, heuristics search algorithms can be applied to find the best possible solutions (probably shortest path). breadth-first search [1] is a well-known algorithm to find the shortest path between a source node and all other nodes in the network. However, a conventional breadth-first search algorithm does not support finding the shortest path between only two nodes. Finding all paths in a single run rather time and space consuming and not useful in a restricted conditional environment like open networks. Another approach to deal with the search problem in a large network is random walks [2] where the path between two nodes can be found by repetitive random paradigm. Using this technique, at each node the next step is taken by selecting one connection (neighbor) of this node randomly. The drawbacks of random walk are basically the repetition that takes place by vising the same node more than once. Furthermore, using this technique the search would take inordinate amount of time and subsequently space. However, a path would be found in some cases faster than heuristic search.Another algorithm to deal with search across an open network is Dijkstra's algorithm [3]. This algorithm does use the same methodology of breadth first search; however, it differs from BFS by not taking into account the length of the connection (edge) between the nodes. It keeps information about each path which had been found before for each node and updates them repeatedly. Similar to BFS, Dijkstra's algorithm is time and space consuming and relatively it has a relatively high complexity (depending on how it is implemented) [4][5]. As shown above all the algorithms which have been described do not consider learning and keep history of pervious less efficient selection. The aim of this paper as will be described in the next sections is to improve the performance of these algorithms. Employing learning and keeping track of previous decisions would make it possible to take better decisions when moving from one node to another during the search process. Furthermore, rearranging the network's topology by adapting the network connections according to the information which have been collected before would improve the search performance in terms of time and accuracy of results. The rest of this paper is organized as follows:Section 2 illustrates general background on overlay network, section 3 related work, then section 4 the system model and algorithms. More after, section 5 deals with the simulation results and analysis. Finally, section 6 contains the conclusions from the applied solutions and suggestions for future improvements.

## 2    Overlay Network

Overlay network is an application network create virtually on top of the physical networks or topologies. Which each node in overlay network initially select its neighbors and built overlay links to minimize its own cost as well as the overall network cost [6]. Overlays network do not suffer the rigidness of physical network, since the logical topology more flexible and increase the adaptive of reconfiguration [7]. The physical topology i.e underlay network can be presented as a graph (G), in the proposed model assume N number of nodes in virtual and physical are equal. The peers communicate with each other and establish self-management and organization over the physical networks. P2P network has a high degree of decentralization compare with other methods [8]. For examples distributed system such as P2P network are overlay network because the P2P nodes run on top of the internet [9]. originally internet build as an overlay network [10], in addition the overlay network creates the foundation of virtual network infrastructure. Fig 1. Shows the structure of overlay network over the internet

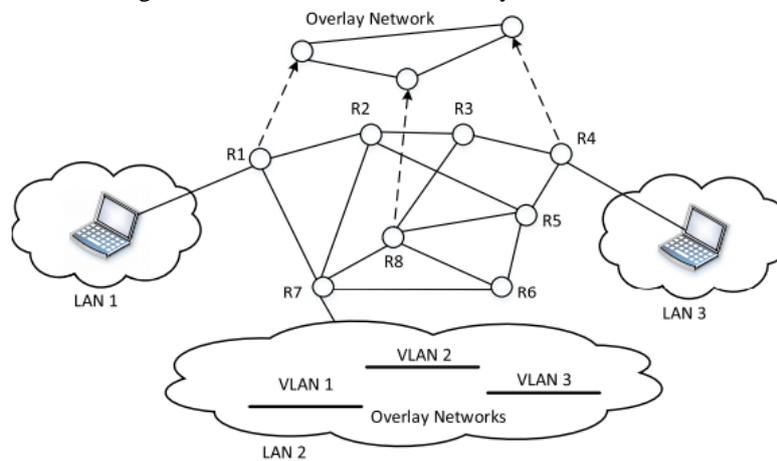

**Fig.1** example of P2P overlay network [9]

## 3.    Related Work

Several methods of self-adaptive overlay network were proposed by the researchers; in this section some approaches will discussed as follow: M. Youssef and the research group in [6] , proposed strategy based on adaptive of overlay network topology using some matrix of traffic estimation, based on node behavior to create overlay link between closest and near-optimal overlay of each node based on heuristic search algorithm,. The topology adaptation leads to change some of traffic because of adding and deleting overlay link in topology. As well as in [7] , the researcher proposed traffic predication model for self –adaptive routing on overlay network topology , which can used a neural network to predicate a traffic link that will used in future. In [8] Peer 2 Peer overlay networking were discussed based on full topology knowledge network, even routing information and the routing manner. The overlay network can provide neighborhoods locations and services applied through directed search, therefore the model can be creating a direct link to the closed manner node in overlay environment. In another work [9 ], a distributed algorithm to solve the problem of optimal content-based routing (OCBR) was proposed based on an assumption, therefore the reconfigured overlay network cannot adapt well to traffic.

## 4.    System Model of Work

To evaluate the fidelity of the suggested solutions the network has been designed as graph with vertices represent nodes and edges represents connections [11]. Each node could be a search initiator or a target of a search message which is initiated by another node. Typically resources in a distributed system would be described using standard description languages or frameworks like WSDL (Web Service Description Language) [12], RDF (Resource Description Framework) [12] and OWL (Web Ontology Language) [13][14][15]. For simulation purposes the resources to be owned by each node, the resources have been described as a set of numbers which are going to be used as a key factor in the matchmaking and adaptation process. Each node is assigned a vector of three values of the form: $r_{i=\langle r0_i, r1_i, r2_i \rangle}$, where each resource element is assigned a value between 0-4 using uniform random distributer function.Furthermore, the matchmaking process between two descriptions, for instance a search request message with a receiver node description is calculated using Manhattan distance measure as follows:

$$D(i.j) = \sum_{k=0}^{k=2} |rk_i - rk_j| \qquad \text{.........(1)}$$

Where:

D(i. j): is the semantic distance between description i and description j.

ri: is the resource description of i.

rj: is the resource description of j.

Accordingly the similarity is computed as the opposite of distance using equation (2) as follows:

$$sim(i.j) = \left( \frac{12 - D(i.j)}{12} \right) 100\% \qquad \text{.................(2)}$$

For example two descriptions of the exactly the same resources are produce a distance =0, which in turn gives a similarity $sim = \left( \frac{12-0}{12} \right) * 100 = 100\%$ while two descriptions with the maximum distance = 12 produce a similarity as follows: $\text{sim}\left(\frac{12-12}{12}\right)100 = 0\%$.

## 5.    Simulation Result and Analysis

The experiment has been designed to study the system performance in terms of search performance and quality of search results. Search performance has been measured by calculating the number of hops to satisfy the search criteria. While the quality of search results is represented by the semantic similarity between the request and node's resource description.The comparisons have been performed between two search algorithms. Namely, standard breadth first search algorithm and proposed guided search algorithm. More after, a performance comparison has been conducted between the two mentioned algorithms with topology adaptation and without topology adaptation.

### 5.1    Simulation Setting

The network has been formulated as a bi-direction graph with vertices represent network nodes and edges represent connections. The level of network subject of this study is on application layer, that means two nodes are connected if they know each other contact addresses.We have used Java to implement the suggested solutions and algorithms. More after, the simulation results have been analyzed and represented using matlab toolkit. Table (1) shows the simulation setting for the first experiment where the network is being modeled as a graph of 200 nodes with each node conned to a maximum of 15 other nodes. More after, a total of 50 request have been created by each node through the simulation time and maximum number of hopes for each node to reach the destination or fail =10.

*Table 1: Simulation Setting*

| Parameter name | Number of nodes | Max connections | Max requests | Number of hops |
|---|---|---|---|---|
| value | 300 | 15 | 50 | 10 |

To study the feasibility of the suggested solutions a set of experiments have been conducted to evaluate the system performance under different conditions. First of a set of experiments have been engineered to study the search performance with a static network setting (i.e. the network is initially created randomly with each node must be connected to at least one other node). Two different search algorithms are implemented and tested under the static network, namely: breadth first search and guided search. The simulation results have shown that guided search is performing better that breadth first search algorithm under some conditions. With smaller network and partially connected nodes no noticeable differences in performance have been noticed. However, with bigger networks and fully connected nodes (each node is connected to the maximum number of nods) guided search performance is improved gradually as the simulation is running. Analysis has shown that nodes are learning from their previous search requests and accordingly forwarding the future requests to the best known connections and so on. More after, the system has been evaluated with network adaptation. Using this technique the node forwards the message to the best contact as in guided search but additionally updates it's list of connections by removing the worst contact ( the connection with resource description that is far away from node description). Updating the node's connections locally would update the whole network globally. The simulation results have shown a gradual increase in system performance in terms of (search success ratio, number of hops to success).

The three experiments have been named as:

Configuration 1: standard search algorithm as config1.

Configuration 2: guided search  as config 2.

Configuration 3: guided search with self-adaptation as config 3.

  As shown in figures (2,3 and 4). The figures are shown that under predefined allowable matching error between 0 (no error) and 12(completely difference description) guided search with self-adaptation has over performed the other two settings. allowable error values in the figures is cut to 6.0 because the system has shown no changes after.

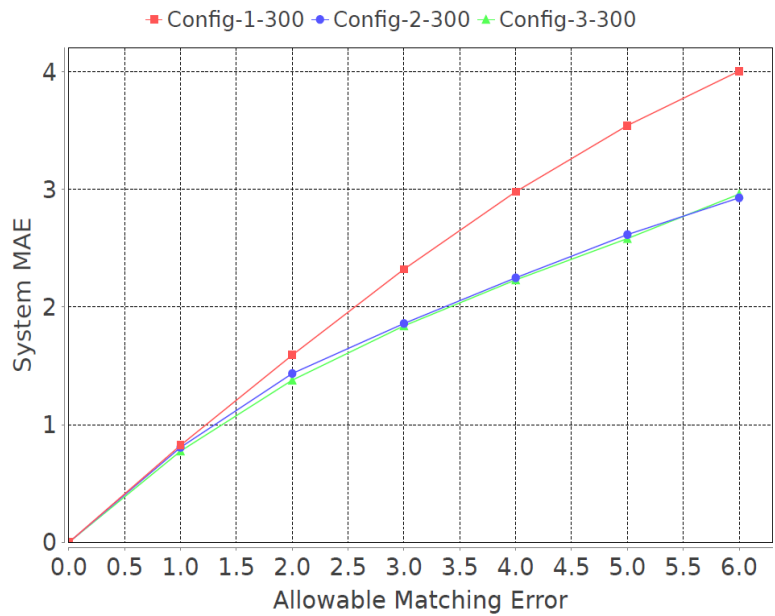

**Fig 2**. A Comparison Between the Three Configurations in Terms of System Mean Average Error

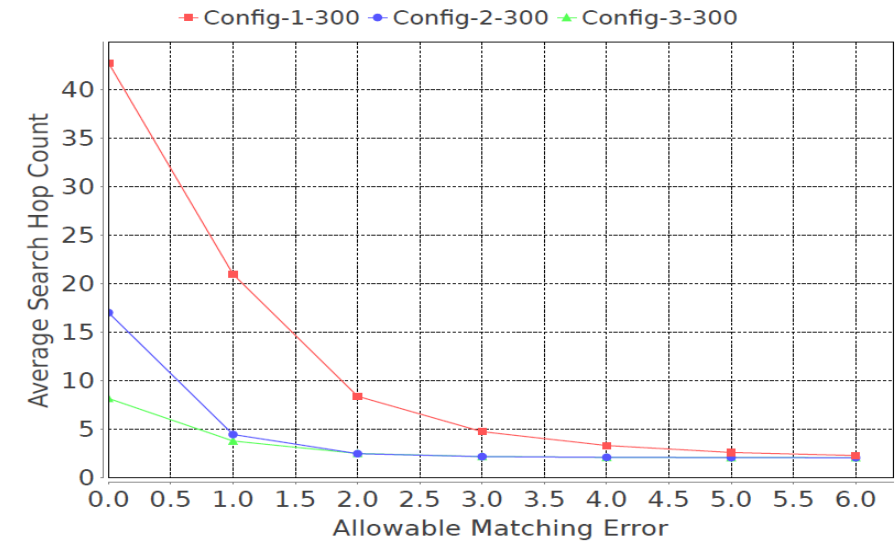

**Fig 3**. A Comparison Between the Three Configurations in Terms of Average Success Hop Count

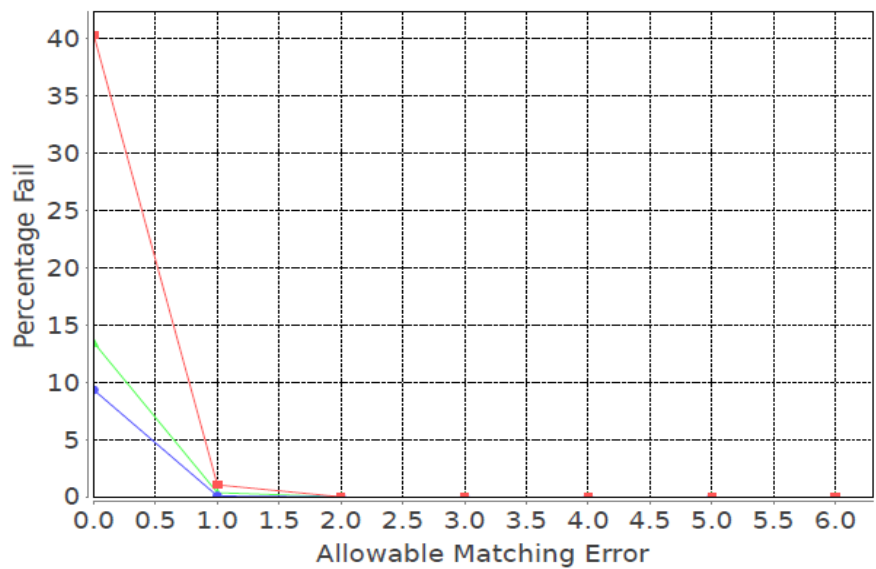

**Fig 4**. A Comparison Between the Three Configurations in Terms of Failure(success) Ratio

The proposed model suggests guided search based on adaptation of network topology based on node's behavior and their own value. Fig 5 shows the initial topology network connection of 50 nodes and topology after adaptation. the figure shows that each node has connections with nodes who is similar or have some of features contain same interests.

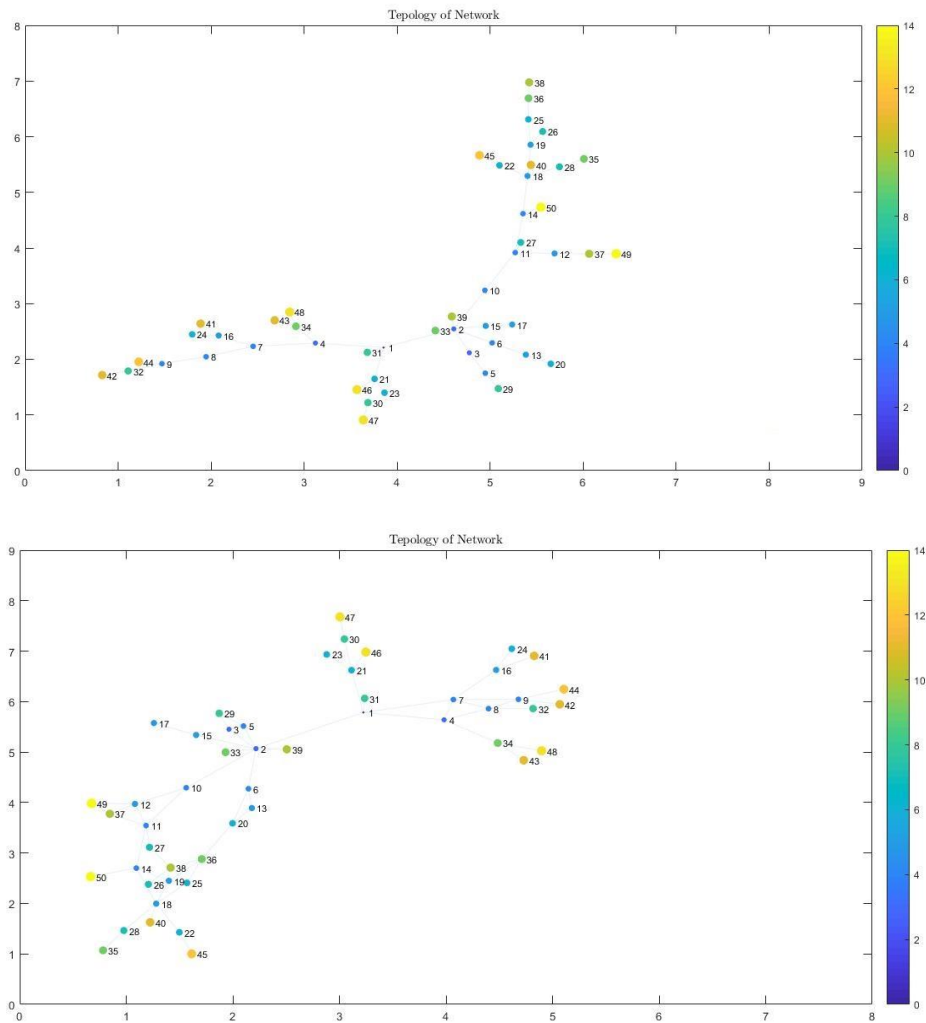

**Fig 5**. (a) initial 50 node network topology   (b) adaptation of the same network

Each node has own features and attributes, based on these values the new connections are established virtually to improve the search performace in overlay network.

## 6.    Conclusion and Future Work

This paper has studied the process of search through a dynamic adaptable network. As the network size increases proportionally, searching the whole network is a time consuming activity. To address this problem a standard breadth first search algorithm has been adapted to support guided search by using the resource owned by each node. Furthermore, the suggested solutions propose the use of the guided search heuristics to self-adapt the network itself which create the phenomena of "local change-global effect". Simulation results have shown that guided search and adaptation are improving the search performance noticeably. Suggestions for future improvements would be by applying the same solutions on more search algorithms and testing them under different network types

## Refrences

[1] Yichun Sun; Xiaodong Yi and Hengzhu Liu, The Communication Analysis of Implementation in Breadth First Search Algorithm, IEEE 2016.

[2] N. Bisnik and A. Abouzeid. Modeling and analysis of random walk search in P2P networks. In Proc. Of Second International Workshop on Hot Topics in Peer-to-Peer Computing (HOT-P2P'05), July 2005.

[3] Jinhao Lu; Chi Dong, Research of shortest path algorithm based on the data structure, IEEE 2012.

[4] Yinglong Xia and Viktor K. Prasanna. TOPOLOGICALLY ADAPTIVE PARALLEL BREADTH-FIRST SEARCH ON MULTICORE PROCESSORS.

[5] Takahiro Tsujimoto, Takuya Shindo, Takayuki Kimura, and Kenya Jin'no . A Relationship between Network Topology and Search Performance of PSO. IEEE World Congress on Computational Intelligence 2012.

[6] Mina Youssef, Balasubramaniam Natarajan, Caterina Scoglio, Adapting the Overlay Network topology based
on Traffic Matrix Estimation , 2007.

[7] Meng Chi, Jianhua Yang,Yabo Liu, and Zhenhui Li , A Traffic Prediction Model for Self-Adapting Routing Overlay Network in Publish/Subscribe System, Mobile Information Systems Volume 2017.

[8] Marcel Waldvogel, Roberto Rinaldi, ACM SIGCOMM Computer Communication ReviewJanuary 2003.

[9] M. Migliavacca and G. Cugola, "Adapting publish-subscribe routing to traffic demands," in *Proceedings of the Inaugural International Conference on Distributed Event-Based Systems (DEBS '07)*, pp. 91–96, ACM, Ontario, Canada, June 2007.

[10]Galluccio L,MorabitoG,PalazzoS,PellegriniM,RendaME,SantiP.Georoy:a location aware enhancement to viceroy peer-to-peer algorithm. ComputNetw 2007;51(8):1998–2014.

[11] Simone A. Ludwig and S. M. S. Reyhani. Semantic approach to service discovery in a grid environment. Journal of Web Semantics, 4:1,13, 2006.

[12] Haluk Demirkan, Robert J. Kau_man, Jamshid A. Vayghan, Hans-Georg Fill, Dimitris Karagiannis, and Paul P. Maglio. Service-oriented technology and management:Perspectives on research and practice for the coming decade. Electron Commer. Rec. Appl., 7:356{376, December 2008.

[13] Eng Keong Lua, Jon Crowcroft, Marcelo Pias, Ravi Sharma, and Steven Lim. A survey and comparison of peer-to-peer overlay network schemes. IEEE Communications Surveys and Tutorials, 7:7293, 2005.

[14] Vassilios V. Dimakopoulos and Evaggelia Pitoura. Performance analysis of distributed search in open agent systems. In Proceedings of the 17th International Symposium on Parallel and Distributed Processing, IPDPS '03, pages 20.2{, Washington, DC, USA, 2003. IEEE Computer Societ

[15] HACINI, Abdelhalim; AMAD, Mourad. A new overlay P2P network for efficient routing in group communication with regular topologies. *International Journal of Grid and Utility Computing*, 2020, 11.1: 30-48.